%Paper: hep-th/9208059
%From: FIORENZO@vana.physto.se
%Date: Mon, 24 Aug 1992 15:50 +0200

\input harvmac.tex
\def\m{\mu}
\def\g{\gamma}
\def\a{\alpha}
\def\n{\nu}
\def\p{\phi}
\def\pa{\partial}
\def\d{\delta}
\def\l{\lambda}
\def\s{\sigma}
\def\t{\tau}
\def\r{\rho}
\def\b{\beta}
\def\ns{\slash \! \! \! \! \nabla }
\def\e{{\rm e}}
\def\sqr#1#2{{\vcenter{\vbox{\hrule height.#2pt
     \hbox{\vrule width.#2pt height#1pt \kern#1pt
           \vrule width.#2pt}
       \hrule height.#2pt}}}}

\Title{\vbox{\baselineskip12pt\hbox{USITP-92-02}\hbox{CERN-TH.6512/92}
}}
{{\vbox{\centerline{ Trace Anomalies from Quantum Mechanics}}}}
\baselineskip=10pt
\vskip -3mm
\centerline{Fiorenzo Bastianelli\footnote{$^\dagger$}{$\!\!\!\! \!\!$
e-mail:
fiorenzo@vana.physto.se}}
\smallskip\centerline
{\it  Institute for Theoretical Physics}
\centerline {\it University of Stockholm}
\centerline {\it Vanadisv\"agen 9 }
\centerline {\it S-113 46 Stockholm, Sweden}
\bigskip

\centerline{Peter van Nieuwenhuizen\footnote{$^*$}{$\! \!\!\!\! \!$
e-mail:
vannieu@cernvm.cern.ch. On leave from ITP, SUNY at Stony Brook, NY 11794,
USA. }} \smallskip\centerline
{\it  CERN, Theory Division}
\centerline {\it CH-1211 Geneva, Switzerland}

\vskip 6mm

\noindent
The  one-loop anomalies of a $d$-dimensional quantum field theory
can be computed by evaluating the trace of the
path integral Jacobian matrix $J$, regulated by an operator
$\exp (- \b{\cal R}) $ and taking the limit $\b$ to zero
\ref\fuji{
K. Fujikawa, Phys. Rev. Lett. 44 (1980) 1733; Phys. Rev. D21 (1980) 2848
and
D23 (1981) 2262.}.
In 1983, Alvarez-Gaum\'e and Witten
\ref\ag{L. Alvarez-Gaum\'e and E. Witten, Nucl. Phys. B234 (1984) 269.}
 made the observation
that one can simplify this evaluation by replacing
the operators which appear  in $J$ and ${\cal R}$
by quantum mechanical operators with the same (anti)commutation
relations.
By rewriting  this quantum mechanical trace  as a path integral
with periodic boundary conditions
at time $t = 0$ and $t = \b$ for a one-dimensional
supersymmetric non-linear sigma model, they obtained the chiral anomalies
for spin ${1\over 2}$ and  ${3\over 2}$ fields and
self-dual antisymmetric tensors in  $d$ dimensions.

Some time ago it occurred to us  that one can also apply these ideas
to the trace anomalies. In a recent paper
\ref\fb{F. Bastianelli, Nucl. Phys. B376 (1992) 113.}
the first author
proposed a bosonic configuration space path integral for a particle
moving
in curved space, and found the corresponding Hamiltonian
${\cal R}$ from the  Schr\"odinger equation.
He exponentiated the factors $\sqrt g$
in the path integral measure by using scalar ghosts, and obtained the
trace anomaly for a scalar field in an external gravitational field in $d=2$.
In this article, we treat the general case of
 trace anomalies for external gravitational
and Yang-Mills fields.
We do  not introduce  a supersymmetric sigma model, but keep  the
original Dirac matrices $\g^\m$ and internal symmetry generators $T^a$
in the path integral.
As a result, we get a matrix-valued  action $S$. Gauge covariance of the
path integral then requires a definition of  the exponential of the action
by time-ordering. The computations are simplified by using
Riemann normal coordinates.
We also replace the scalar ghosts by vector ghosts
 in order to exhibit the cancellation of all
 divergences at finite $\b$ more clearly.
Finally we compute the trace anomalies
in $d=2$ and $d=4$.
\vskip 3mm

\hbox{ CERN-TH.6512/92}

\hbox{ May 1992}

%\Date{3/92}
%\draftmode
\vfill\eject
\baselineskip=16pt
\footline={\hss\tenrm\folio\hss}
\pageno=1

\newsec{Introduction}
The one-loop anomalies in local symmetries of a $d$-dimensional quantum
field theory can be computed, as first demonstrated by Fujikawa \fuji,
by evaluating the trace of the Jacobian matrix $J$ that arises when one
varies  the integration variables of the path integral under this local
symmetry. This trace has to be regulated  because the Jacobian
is an infinite dimensional matrix, and  as regulator one may take an
operator
 of the form $\exp ( - \b {\cal R})$. The anomaly is then obtained by
letting
$\b$ tend to zero. In general, there will be  divergent terms proportional
to powers of $\b^{-1}$, and a finite remainder. The latter is the anomaly,
while the former are discarded by using several regulators
whose $\b$'s satisfy suitable relations as in
Pauli-Villars regularization. The justification for discarding the
divergences in anomalies is that the effective action is made finite by
adding suitable counterterms so that its variation should also be finite.
(If one regulates the quantum field theory with dimensional
 regularization,
the counterterms which make the one-loop  effective action finite
produce also  anomalies \ref\duff{M. Duff, Nucl. Phys. B125 (1977)
334.}, which are finite since the one-loop pole is cancelled
by a zero due to classical gauge invariance).

The choice of a regulator ${\cal R}$ is important.  For some cases one
knows from computations a regulator which yields consistent anomalies,
or a regulator which yields a covariant anomaly, but only for
consistent  anomalies has a general theory  been constructed
\ref\pvn{A. Diaz, W. Troost, A. Van Proeyen and P. van Nieuwenhuizen,
Int. J. Mod. Phys. A4 (1989) 3959\semi
M. Hatsuda, W. Troost, A. Van Proeyen and P. van Nieuwenhuizen,
Nucl. Phys. B335 (1990) 166\semi
F. Bastianelli, A. Van Proeyen and P. van Nieuwenhuizen,
 Phys. Lett. B253 (1991) 67. }.
This  theory provides  a method to construct
regulators which: $i)$ preserve certain local symmetries at the quantum
level, and $ii)$ produce anomalies in other symmetries which
satisfy the consistency conditions.
For Einstein symmetries (general coordinate invariance), Fujikawa
already determined the  regulators which satisfy $(i)$.
For covariant anomalies, no general theory is known as yet, although in
practice one  usually can guess what these covariant regulators will be.
In  our work below, we are in the happy circumstance that the consistent
regulators which preserve the Einstein symmetry are actually covariant, as
we shall discuss.

The local symmetries we shall consider are local scale symmetries.
The corresponding anomalies are called Weyl anomalies, or trace anomalies
because they arise  as a non-vanishing trace of the stress tensor
$T_{\m\n}$ at the quantum level. For example, consider a real scalar
 field
$\p$, which couples to gravity in $d$ dimensions with action
\eqn\osca{S = \int dx {\sqrt g}\
{1\over 2} \biggl \lbrack
g^{\m\n} \pa_\m   \p \pa_\n  \p - \xi R  \p^2
\biggr \rbrack,}
where $\xi = { d-2 \over {4(d-1)}}$.
We shall always work in Euclidean space so that $g ={\rm det}\ g_{\m\n}$.
This classical action is invariant under the following Weyl rescalings
\eqn\ows{g_{\m\n} \to {g_{\m\n}}' = \Omega g_{\m\n} ; \ \ \ \ \ \ \
\p \to {\p}' = \Omega^{{
{2-d} \over 4}} \p .}
The general theory of consistent regulators states that a consistent
regulator without Einstein anomalies is obtained by using  $\tilde
\p = g^{1\over 4}  \p $ as quantum field. Clearly
\eqn\owf{\delta_{_W} \tilde \p = {1\over 2} \s \tilde \p, }
where $\s = \log \Omega $ is taken to be infinitesimal.
Rewriting the action in \osca\ in term of $\tilde \p$, the following
regulator is obtained
\eqn\oregul{ {\cal R}^{cons} =-  g^{-{1\over 4}} \pa_\m
g^{{1\over 2}}   g^{\m\n} \pa_\n g^{-{1\over 4}} - \xi R .}
This is clearly not an Einstein scalar. Under a rescaling  of the external
metric in the path integral $ Z[g] = \int {\cal D} \tilde \p
\exp ( - \hbar^{-1} S[g,\tilde\p])$, followed by a compensating rescaling
of $\tilde \p$  as in \owf\ such that $S$ remains invariant, the
following regulated Jacobian is produced:
\eqn\regjac{
 \int dx {\sqrt g}\ \s(x)  g^{\m\n}(x)
\langle T_{\m\n} \rangle
=  2 \hbar \lim_{\b \to 0} \biggl ( {\pa \d_{_W}
\tilde \phi \over { \pa \tilde \p}} \exp (-\b {\cal R}^{cons}) \biggr), }
where we have defined the stress tensor by
\eqn\ostress{ T_{\m\n} ={2\over{\sqrt g}} {\d S \over {\d g^{\m\n}}} .}
Since the Jacobian contains no derivatives, we can use the cyclicity
of the regulated trace \ref\anna{ A. Ceresole, P. Pizzocchero and P.
van Nieuwenhuizen, Phys. Rev. D39 (1989) 1567. }
 to cycle the factor
$g^{-{1\over 4}}$ on the right-hand side of ${\cal R}^{cons}$ in \oregul\
to the left-hand side, obtaining the  trace anomaly
of a scalar field in $d$ dimensions
\eqn\oasca{ \eqalign{ &A_0^{(d)}
 = \int dx {\sqrt g}\ \s(x) \langle T \rangle
=  \hbar \lim_{\b \to 0}   {\rm Tr}\ \s(x) \exp (-\b {\cal R}^{cov}_0)
\cr &{\cal R}^{cov}_0 = -g^{-{1\over 2}} \pa_\m  g^{1\over 2} g^{\m\n}
\pa_\n -\xi  R = -\nabla^2 -\xi  R ,\cr}}
where the subscript zero  indicates that
we are dealing with spin zero fields and
where $T$ denotes the trace of the stress tensor.
For Einstein anomalies, one cannot replace ${\cal R}^{cons}$
by ${\cal R}^{cov}$ as the Jacobian contains derivatives \anna.
The reason that
this is possible  for Weyl symmetries follows already from the
well-known fact that the consistent Weyl
anomaly is Einstein-covariant.

The problem is now to evaluate this trace. Fujikawa \fuji\
and others \anna\ have used a complete set of plane-waves and
integrated over plane-wave momenta, but the expressions become
extremely unwieldy. DeWitt \ref\dewitt{B.S. DeWitt, in ``Relativity,
Group and Topology'', eds. B.S. DeWitt and C. DeWitt (Gordon Breach,
New York, 1964); ``Relativity, Group
and Topology II'' eds. B.S. DeWitt and R. Stora (North Holland,
Amsterdam, 1984).}
did develop, much
earlier, a heat kernel analysis, according to which an ansatz for the
answer is made, which is then verified by explicit computation
order by order in $\b$. In  the ansatz, ``bilocal'' tensors appeared
(such as the Synge world function $\s(x,y)$, which measures the
square of the distance between two points)
and a calculus for the left and/or right derivatives of such objects was
needed.
Of course this approach yields  much more than only
the anomalies, for example one obtains in this way the complete one-loop
effective action. Yet, a simpler method, which may  only give the anomalies,
might be welcome.
Some time ago, Alvarez-Gaum\'e and Witten \ag\ observed that one can in
general compute anomalies in a much simpler way by replacing the
$d$-dimensional operators  in $J$ and ${\cal R}$, such as $x^\m$,
${\pa \over {\pa x^\m}}$, the Dirac matrices $\g^\m$, the metric
$g_{\m\n}(x)$, the connection $A_\m(x)$ and the corresponding generator
of internal symmetries $(T^a)_i{}^j$, by quantum mechanical objects
$ q^\m(t)$ and $p_\m(t) $, $ \psi^\m(t),\  g_{\m\n}(q),\ A_\m(q) $
and $ c^{*i}(t) (T^a)_i{}^j c_j (t)$, where $ c^{*i} $ and
$ c_j $  are anticommuting ``ghosts'' which contract the indices of
internal symmetry generators such as Yang-Mills or Lorentz matrices.
All that is needed is that  they satisfy the same anticommutation rules.
They then considered chiral anomalies and rewrote the trace of the
quantum mechanical operator $ J \exp (-\b {\cal R})$,
with $J=\g^5$ as a one-dimensional path integral
\eqn\opbc{ Z =\int_{^{PBC\ {\rm for}\ q, \psi}_{ABC\ {\rm for}\ c,c^*}}
 {\cal D} q^\m {\cal D} \psi^\m
 {\cal D} c_i {\cal D} c^{*i}\  \exp \biggl (- \int_0^\b
dt \ {\cal L}^{(\s)} \biggr ), }
where ${\cal L}^{(\s)} $ is the Lagrangian for a one-dimensional
supersymmetric non-linear sigma model. The $d$-dimensional chirality
operator $\g^5$ became the fermion number operator $(-)^F$,
because $(-)^F$ anticommutes with $\psi^\m$ just as  $\g^5$ with  $\g^\m$.
Since
$(-)^F$ commutes with $q^\m$ and the ghosts,
the $\psi^\m$ fermions acquired periodic
boundary conditions (PBC), whereas the $c^{*i}$ and $c_j$ kept their
antiperiodic boundary conditions (ABC).
A simplification was due to the
topological nature of the chiral anomaly:
the path integral itself  is $\b$-independent and hence  could be
evaluated
by taking $\b$ small, and keeping only non-vanishing terms. In particular,
no divergences proportional to $\b^{-1}$ could occur.
For the actual evaluation,
Alvarez-Gaum\'e and Witten expanded $q^\m(t)$ and $\psi^\m(t)$
in fluctuations about  a classical constant solution $q^\m(t) = q^\m_0$
and $\psi^\m(t)= \psi^\m_0$, and only the terms quadratic in
fluctuations did contribute. The result for the anomaly could be written
in a covariant form by using normal coordinates and,  since they used
a covariant regulator,
this covariantly looking form was also valid in general coordinates.

In this article we shall consider the trace anomalies, and extend the
work of Alvarez-Gaum\'e and Witten to this case.
There are, however, important differences:

i) The trace anomaly has no topological meaning,
and as a consequence its path integral representation
does depend on the regulating parameter
$\b$. We must therefore expand in terms of $\b$, which amounts to a
loop
expansion on the world line with coordinate $t$. For the one-loop trace
anomaly in $d$ dimensions we shall need ${d\over 2}+1$
 loops on the world line.
Hence the algebra becomes more complicated than for  chiral anomalies
and subtle problems which did not need to be addressed in \ag , now have to
be solved. For example, the na\"{\i}ve relation between
the Hamiltonian (=regulator)
and the corresponding Euclidean action in the path-integral breaks down,
and extra terms proportional to the curvature appear. The occurrence of such
terms was first noticed by DeWitt \dewitt . (In fact, in his well-known 1948
article \ref\feynman{R.P. Feynman, Rev. Mod. Phys. 20 (1948) 367.},
Feynman had
already indicated that for models such as non-linear sigma
models, it
might not be correct to use the classical action and classical paths
of a free particle to define the path integral.)

ii) For the case of  spin $1\over2$ fields, our
Jacobian has no factor $\g^5$, hence no operator
$(-)^F$; therefore
all fermions will now acquire antiperiodic boundary conditions. Hence,
there are no constant modes $\psi^\m_0$  about which to expand.
This, in itself, is no problem: we could expand
$\psi^\m$  only into antiperiodic fluctuations.

iii)  As pointed out in \ag,
the anticommuting creation and absorption operators $c^{*i}$
and $c_i$ should only act on the one-particle states in the
quantum-mechanical trace of the operator $J \exp (- \b {\cal R})$
in order that  $c^{*i}T^a{}_i{}^j c_j $ acts in the quantum-mechanical
problem in the same way as the matrices  $T^a$
did in the original problem. Thus we would need a quantum-mechanical
projection operator, which should become an extra internal kernel
in the corresponding one-dimensional path integral. Although
we have found this kernel,
see (B.8), the boundary conditions to which it corresponds
lead to complicated propagators, and therefore we decided to follow a
different path. We keep at all times  the Dirac matrices $\g^\m$
and  internal symmetry generators $T^a$
as matrices, and never introduce any $\psi^\m(t),\ c^{*i}(t)$
or $ c_i(t)$ at all.
 Although to our regret supersymmetry  leaves the stage  by this approach
(it could have been present but for the spin $1\over 2$ regulator, and it
would anyhow have been broken by the boundary conditions),
it seems the simplest way to compute
the path integral.  In fact, we shall rewrite  the action
$S$ in terms of a connection $A_\m$ and a potential  $V$, which may
depend in any way on the matrices $\g^\m$ and $T^a$. Note that in [2]
this same approach was followed for  $c^{*i}$ and $c_{i}$, but not for
the Dirac matrices $\g^{\m}$, which were replaced by $\psi^{\m}(t)$.
As a result the authors of ref. [2] obtained a supersymmetric
sigma model,
and the zero mode $\psi_{0}$ was useful
to write the chiral anomaly in terms of
p-forms in $d$ dimensions. Since p-forms do not occur in trace anomalies,
there is no advantage in replacing $\g^{\m}$ by $\psi^{\m}(t)$ in our
case.

iv) If one defines the path integral by discretizing time into
$ t_{_0} =0,t_{_1}, \dots, t_{_N},t_{_{N+1}}=\b$, the
measure
for $q^\m(t)$
contains a factor
$\sqrt g$, which is due to integrating out the momenta $p_\m(t)$
in the path integral. This factor is well-known from non-linear sigma models,
and it is usually exponentiated, yielding a term with $\d (0)$
which cancels the leading-loop divergences of the non-linear sigma model
\ref\lee{ E.S. Abers and B.W. Lee,  Phys. Rep. 9 (1973) 1.}.
In \fb\ these factors were exponentiated by using scalar ghosts
\eqn\osgh{{\cal L}_{gh}= b(t) {\sqrt {g (q(t))}} c(t)  .}
Below we shall instead use vector ghosts
\eqn\ovgh{ {\cal L}_{gh} = {1\over 2} g^{\m\n}(q(t))
\biggl ( b^\m(t) c^\n(t)
+ a^\m(t) a^\n(t)\biggr ) ,}
which give the same result, but which simplify the bookkeeping of loops
because the $\dot q \dot q$ part of the action is of the same form
\eqn\oq{ {\cal L} = {1\over 2} g^{\m\n}(q(t)) \biggl ( \dot q^\m(t)
\dot q^\n(t) \biggr ) .}
All divergences between the  various
non-ghost and ghost loops cancel for finite $\b$,
which is rather remarkable
as non-linear sigma models contain interactions with
derivatives, confirming the measure for $q^{\m}(t)$.

It is well-known that if one has $N$ intermediate points $t_{_1},t_{_2},
 \cdots ,
t_{_N}$ between the initial and final times $t_i$ and $t_f$, one needs
$(N+1)$ intermediate sets of momentum states, leading to $(N+1)$ factors
$\sqrt{{\rm det} \ g}$ \ref\schul{L.S. Schulman,
\lq\lq Techniques and applications of path integration\rq\rq,
(John Wiley and Sons, New York 1981).}.
We shall exponentiate only $N$ intermediate
factors $\sqrt{g}$. As a result $b^{\m}$ and $c^{\m}$
 vanish at the boundaries,
and we can expand them into sin$(n\pi t/ \b)$. The
extra factor $\sqrt{g}$ appears
then in the completeness relation as
$ \int dq\sqrt{g(q)} |q\! >< \!q|= \hat 1 $, and
leads to the transition amplitude \fb, which is
used to evolve wave functions
\eqn\opvn{ \psi(q^{\m}_{f},t_f)=\int d q_i \sqrt{g(q_i)}< \!q^{\m}_f,t_f|
q^{\m}_i, t_i\! >
\psi(q^{\m}_i,t_i)}

v) In order to determine which Euclidean action corresponds to a given
Hamiltonian (=regulator), there are in principle two ways to proceed. Either
one starts with the Hamiltonian operator formalism,
 discretizes time, introduces
phase-space variables, integrates out the momenta, and finally transforms
from the discrete-time basis to, for example,
a trigonometric basis. Or one bypasses this
cumbersome but fundamental approach, and determines directly from the
Schr\"odinger equation which Hamiltonian operator belongs to a given
path integral. In the latter approach one \underbar{starts} thus with the
Euclidean action and the path integral on a trigonometric basis, computes it
to the required order in $\b$, and then finally finds the Hamiltonian to
which it corresponds.
We shall take a very general Euclidean action so that we can reach all
the Hamiltonians we are interested in. The main problems with the former
approach (the Hamiltonian operator approach) are the following.
a) \underbar{Given}
our regulator, there are \underbar{no ordering ambiguities} in the
Hamiltonian itself; however, one has to evaluate the matrix elements
$< \!x|\exp (- \epsilon H)|p\! >$,
and here we should first expand the exponent, then
move all $\hat x$ to the left and all $\hat p$ to the right, and then
re-exponentiate the $c$-number result. For the
harmonic oscillator this program
is easily
 executed,
but for non-linear sigma models it seems complicated. b) One has to
compute the exact Jacobian for the transition from a discrete-time basis
[the variables $q(t_j)$] to a trigonometric basis
 [the coefficients $q_n$ in
$q(t)=\sum q_n {\rm sin}(n\pi t/\b) $]. Usually, one first guesses how the
continuum action will look like, and then introduces a trigonometric basis
whose measure one fixes such that it gives
the correct result for the harmonic
oscillator. Clearly, it would be better if no
 guesses or approximations were
made, but we found this problem too difficult to solve. For further
details, see appendix B.
We shall thus follow
Feynman's original method of deriving the Schr\"odinger equation, but not use
the action which na\"{\i}vely corresponds  to the regulator, but rather a quite
general action, which leaves room for extra terms.

vi) The action we consider is of the form
\eqn\owe{ S = \int_0^\b dt \ ({\cal L} + {\cal L}_{gh} +
A_\m \dot q^\m +V ).}
The objects $A_\m$ and $V$ are  matrices, and  in order to define the
path integral
\eqn\opi{Z =\int {\cal D}q{\cal D}b{\cal D}c{\cal D}a\
 \exp ( -S),}
we must decide how to order the various terms in the expansion of $\exp (-S)$.
If one uses a definition of the configuration
space path integral where time is
discretized, it is well-known that gauge invariance fixes an ambiguity with
respect to the point at which we should take the gauge field. (The midpoint
rule fixes the \lq\lq Ito-ambiguity\rq\rq, see \schul .)
 We shall expand all
fields on a given trigonometric basis, and hence no ambiguity seems present.
 Yet, for a matrix-valued action, there is an ambiguity on
 how to order the terms in the expansion of $\exp (-S)$.
We shall show that gauge covariance (not invariance)
of our matrix-valued path integral
 selects time-ordering.
 As a mathematical exercise, we could have considered  path integrals
without time-ordering, but, as already said, this would violate gauge
invariance;  also, from a canonical operator point of view, time-ordering
seems natural.

In the following sections we shall describe these issues in detail.
In sect. 2 we  derive the regulators of the classical Weyl-invariant
field theories,
of which we shall compute the  trace anomalies.
In sect. 3 we define the path integral with matrix-valued connections and
potentials. In sect. 4  we first expand the metric, connection and potential
in terms of Riemann normal coordinates, and expand the quantum fluctuations
of the coordinate and vector ghost fields in a Fourier series.
We then evaluate
the path integrals by making an expansion in terms of the parameter $\b$.
We find the propagators and compute connected as well as
disconnected Feynman diagrams on the world line. The result is a general
formula for the trace anomalies in any
dimension for abstract matrices $A_{\m}$
and $V$ in (4.22). Finally, in sect. 5,
we apply this formula to the specific case of spin
$0,{1\over 2}$, and $1$
in $d=2$ and $d=4$, coupled to external gravitational and Yang-Mills fields.
In sect. 6 we present our conclusions and comments.

\newsec{The regulators for classical Weyl-invariant theories}

In this section we obtain the regulators for the fields in the dimensions
we are interested in. Before starting the discussion,
we present our conventions.
The Riemann curvature is defined by
\eqn\pvnrc{[\nabla_{\m},\nabla_{\n}]V^{\r}=R_{\m\n}{}^{\r}{}_{\s}V^{\s},}
where $V^{\m}$ is an arbitrary vector and $\nabla_{\m}$ the usual covariant
derivative, which commutes with the metric. The Ricci tensor is given by
$R_{\m\s}=R_{\m\n}{}^{\n}{}_{\s}$ and the scalar curvature by
$R=g^{\m\s}R_{\m\s}$.

The action of a scalar field coupled to gravity
and gauge fields reads as follows
\eqn\as{S_{0} = \int dx {\sqrt g}\
{1\over 2} \biggl \lbrack
g^{\m\n} (\pa_\m +A_\m) \bar \p (\pa_\n + A_\n) \p - \xi R \bar \p \p
\biggr \rbrack,}
where $\xi = { d-2 \over {4(d-1)}}$.
The field $\p$ transforms according to a given representation
of the gauge group and $\bar \p$
according to the conjugate one, so that
$\bar \p \p$ is a scalar under gauge transformations.
The latter can be parametrized as follows
\eqn\gt{A_\m \to {A_\m}' = U \pa_\m U^{-1} + U A_\m U^{-1} ;
\ \ \ \
\p \to {\p}' = U \p ,}
where it is understood that on the scalar field $\phi$
 the group
element $U$ acts in the
corresponding representation (whose indices are
suppressed).
Under  Weyl transformations
\eqn\ws{g_{\m\n} \to {g_{\m\n}}' = \Omega g_{\m\n}; \ \
 A_\m \to {A_\m}' = A_\m;
\ \ \p \to {\p}' = \Omega^{{ {2-d} \over 4}} \p }
the Ricci scalar transforms as
\eqn\R{R \to R' = \Omega^{-1} \biggl \lbrack
 R +(d-1) \nabla^\m \pa_\m \log
\Omega  + {1 \over 4}(d-1)(d-2)g^{\m\n} (\pa_\m \log \Omega )
(\pa_\n \log \Omega)\biggr \rbrack; }
 as a consequence the action \as\ is invariant.
The choice of ${\cal R}$ is not arbitrary if one wants to obtain
a consistent anomaly \ref\zum{W.A. Bardeen and B. Zumino,
 Nucl. Phys.  B244 (1984) 421.}.
A general method
to identify such a regulator is to appeal to
a Pauli-Villars regularization, which guarantees the consistency
of the anomaly. For further details on this procedure, we refer
directly to \pvn.
For our purposes, we use
the Fujikawa variables  $ {\tilde \p } \equiv g^{1\over 4} \p $,
 so that
one is assured that there will be no
gravitational anomalies, and
we obtain the following expression
for the regulator
\eqn\resca{
{\cal R}_\p = -\nabla^2_A - \xi R. } Here the gauge-covariant
Laplacian $  \nabla^2_A $ acts in the representation which the
field $\p$ belongs to. If $\phi$ is real, then the anomaly is given
by \oasca, otherwise it is twice as large.

The Weyl coupling of a complex (Dirac) spin ${1\over 2}$ field
$\psi$ is described by the action
\eqn\af{\eqalign{&S_{1\over 2} = \int dx e\  \bar \psi
\ns \psi, \cr
&\ns =e_m{}^\m  \gamma^m \biggl (
\pa_\m + \omega_\m +A_\m \biggl ), \cr}}
where $e_\m{}^m $ is a vielbein for the metric $g_{\m\n}$,
 $e_m{}^\m$ is its
inverse, $e={\rm det} \ e_\m{}^m$ and $\gamma^m$ are the
$SO(d)$ Dirac matrices.
The spin connection $ \omega_\m =
{1\over 4}\omega_{\m mn} \gamma^{mn}$, with
$\gamma^{mn} \equiv {1\over 2} \lbrack \gamma^m, \gamma^n \rbrack$,
takes values in the spinorial representation of the
$SO(d)$ Lie algebra
and it is a function of the vielbein,
  while the gauge connection $A_\m = A^a_\m T_a$
takes values in an arbitrary representation of an arbitrary Lie group.
The Weyl transformations now read  as follows
 \eqn\wf{e_\m{}^m \to {e_\m{}^m}' = \Omega^{1\over 2}e_\m{}^m ;
  \ \  \psi \to {\psi}' = \Omega^{{
{1-d} \over 4}} \psi. }
Contrary to the scalar case, there is no need to add a
curvature-dependent term
to achieve invariance.
 To obtain the spin $1 \over 2$  regulator, we use again the fields
$\tilde \psi = g^{1 \over 4}\psi$, and find
\eqn\dirac{ \eqalign{
{\cal R}_\psi &=-
 \ns \  \ns = - \nabla^2_{A,\omega} -{1\over 4} R - {1\over 2} F_{\m\n}
\g^{\m\n} \cr
A_{1\over 2}^{(d)}
& = -\hbar \lim_{\b \to 0} {\rm Tr} \biggl ( \s
e^{ - \b {\cal R}_\psi } \biggr ).
\cr}}
For a real (Majorana) fermion one gets half this result.

 Finally we consider the spin 1 case,
which is classically Weyl-invariant only in $d=4$.
We use the background-field method by splitting the gauge field
into a background part $A_\mu$ and a quantum part $A_\m^{(qu)}$.
 Adding a gravitationally  and background covariant
gauge-fixing condition $\nabla^{\m}_AA_\m^{(qu)} $
suitably weighted, one finds
\eqn\agfgf{S_{1}^{gf} = \int d^4x {\sqrt g}\
 {\rm Tr} \biggl \lbrack {1\over 4}
g^{\m\l} g^{\n\s} F_{\m\n} F_{\l\s}
+ \Pi \biggl ( \nabla^\m_A A_\m^{(qu)} - {1\over 2} \Pi
\biggr )  + B \nabla^\m_A
 (\nabla^A_\m C +
[ A_\m^{(qu)} , C ] )
\biggr \rbrack,}
where the subscript A indicates that the derivatives are
background-gauge-covariant.
 We contract $A_\m^{(qu)}$ with a vielbein field to obtain a tangent
space index, and upon
 elimination of the auxiliary field $\Pi$, we find for the regulator of the
spin 1 field:
\eqn\pvnram{R_{A}=-\delta_m{}^n
\nabla^{2}_{A,\omega}- R_m{}^n - 2 F_m{}^n.}
In addition we need the regulator for the ghosts. Both $B$ and $C$ are
to be considered as real, and their regulator is
\eqn\pvnrbc{R_{BC}= -\nabla^{2}_A.}
These regulators  act in the adjoint
representation, thus
the gauge connection as well as the $ F_{mn}$  term in eq. \pvnram\
are absent for Abelian gauge fields.
The trace anomaly is  given by
\eqn\trspo{
A_{1}^{(4)}
= \hbar \lim_{\b \to 0} {\rm Tr} \biggl ( \s
e^{ -\b {\cal R}_A }\biggr )
- 2 \hbar \lim_{\b \to 0} {\rm Tr} \biggl ( \s
e^{ -\b {\cal R}_{BC} }\biggr ),}
where we used $\delta \tilde A_\m^{(qu)} = {1\over 2} \sigma
\tilde A_\m^{(qu)}$,
 $\delta \tilde C =  \sigma \tilde C$, and
 $\delta \tilde B = 0$,
with $\tilde A_\m^{(qu)} = g^{1\over 4}  e_m{}^\m A_\m^{(qu)}$,
 $\tilde C = g^{1\over 4} C$, and $\tilde B = g^{1\over 4} B$.
Perhaps we should  note here that the
 gauge-fixing term in \agfgf\ na\"{\i}vely destroys  the Weyl invariance.
In fact, with the following natural definition of
 the Weyl transformations
 on the $\Pi,B,C$ fields,
\eqn\2{\eqalign{ &\Pi \to \Pi' = \Omega^{-1} \Pi \cr
& B \to  B' = \Omega^{-1} B \cr
& C \to C' = C ,\cr}}
the action is not invariant and there is no way of fixing the above
transformation rules to achieve invariance.
Under an infinitesimal Weyl transformation with $\Omega = e^\s =
1+ \s+\dots$, the action changes by
\eqn\change{\eqalign{\delta S_1^{gf} &= \int d^4x {\sqrt g}\
{\rm Tr} \biggl \lbrack \Pi (g^{\m\n} \pa_\n \s) A_\m^{(qu)} +
B (g^{\m\n} \pa_\n \s) (\nabla^A_\m C +
[ A_\m^{(qu)} , C ] )\biggr \rbrack \cr
&=
\delta_{^{BRST}} \int d^4x {\sqrt g}\  {\rm Tr}  \biggl \lbrack B
(g^{\m\n} \pa_\n \s)
A_\m^{(qu)} \biggr \rbrack\cr}}
In
 the last line  we have made use
of the anticommuting BRST operator defined by
\eqn\BRST{ \eqalign{&\delta_{^{BRST}} A_\m^{(qu)} =
\nabla^A_\m C +[A_\m^{(qu)}
 , C] \cr
&\delta_{^{BRST}} C = -{1\over 2} \{ C,C \} \cr
&\delta_{^{BRST}} B = \Pi \cr
&\delta_{^{BRST}} \Pi = 0 ,\cr}}
which shows that  the Weyl variation of the action is BRST-exact.
For this reason it has zero expectation value between physical
states  \ref\nil{N.K. Nielsen and P. van
Nieuwenhuizen, Phys. Rev. D38 (1988) 318.}.

\newsec{Quantum mechanics with matrix-valued potentials in curved spaces}
We want to find now a path integral representation of the traces
which yield  the Weyl anomalies. The typical trace
we have to compute  has the following structure
\eqn\ty{\eqalign{
&I= \lim_{\b \to 0}
{\rm Tr }\  {\rm e}^{-\beta H} \cr
&H= -  (\nabla^\m + A^\m )(\pa_\m + A_\m) + S,\cr}}
where $H$ acts on fields which are scalars under reparametrization
and transform in a fixed, but otherwise unspecified,
representation of the gauge group. The matrices $A_{\m}$ and $S$ transform
as vectors and scalars under reparametrizations, respectively, but are
furthermore quite general.
Our strategy will be as follows. We first rewrite the trace in \ty\
using the operatorial formalism of quantum mechanics.
Then we employ the path integral formulation of
quantum mechanics to explicitly compute the trace.
As is well-known, one can derive the path integral using
the operatorial formulation of quantum mechanics and vice versa.
As explained in the introduction, it is easier to start with a proper
definition of the path integral
on curved spaces, from which we derive unambiguously
the Hamiltonian operator.

The trace in \ty\ can be taken over the Hilbert space generated
by the following basis vectors:
\eqn\basis{ \vert x\! > \otimes \vert i\! >
 \equiv \vert x,i \! > }
where the kets $\vert i\! >$ correspond to a basis in the
representation space  of the gauge group
and $ \vert x\! > $
are the usual kets of configuration space.
The scalar product is defined by
\eqn\sp{< \! x,i\vert y,j\! > = {\delta(x-y) \over{ \sqrt {g(x)}}}
\delta_i{}^j }
and the resolution of the identity takes the following form
\eqn\ide{\hat 1 = \sum_i \int d x {\sqrt g}\
\vert x,i \! > < \! x,i\vert}
Any state $ \vert \psi\! > $ and  operator $\hat G$  have  the
following components in the above basis
\eqn\com{\eqalign{&\psi_i(x) = < \! x,i \vert \psi \! > \cr
&G_i{}^j (x,y) = < \! x,i\vert \hat G \vert y,j\! > \cr}}
In particular, the usual Hermitian position operator $ \hat q^\m$,
 Hermitian momentum operator $ \hat p_\m$ and  antiHermitian
Lie algebra generator $\hat T^a$ have the following
matrix elements
\eqn\matel{\eqalign{
&< \! x,i \vert \hat q^\m \vert y,j \! > =
 x^\m {\delta(x-y) \over {\sqrt {g(x)}}} \delta_i{}^j \cr
&< \! x,i \vert \hat p_\m \vert y,j \! > =
-i \hbar g^{-{1\over 4}}(x) {\pa \over {\pa x^\m}}
 g^{1\over 4}(x)
{\delta(x-y) \over {\sqrt {g(x)}}} \delta_i{}^j \cr
&< \! x,i \vert \hat T^a \vert y,j \! > = T^a{}_i{}^j
{\delta (x-y) \over {\sqrt {g(x)}}} .\cr}}
One may check that $\hat p_\m$ is represented by $-i\hbar g^{-{1 \over 4}}
\pa_\m g^{1 \over 4}$ on wave functions $\psi_i(x)$, and
that it is Hermitian with respect to
 the inner product $ < \! \phi | \psi \! > =\int dx g^{1 \over 2}
\phi^* \psi$.

A gauge transformation
is implemented by a unitary operator $\hat U$
and we have  the following transformation properties for
 the  wave function and the  transition amplitude
\eqn\trprop{\eqalign{
&\psi_i(x) = < \! x,i \vert \psi \! > \to
< \! x,i \vert \hat U \vert \psi \! > = U_i{}^j(x)
\psi_j(x) =\psi_i(x)' \cr
&< \! x,i\vert \e^{- \beta  \hat H} \vert y,j\! >
\to  U_i{}^k(x) < \! x,k\vert \e^{- \beta  \hat H}
 \vert y,l\! > (U^{-1})_l{}^j(y), \cr}}
where the Hamiltonian operator has the following form
\eqn\Ham{ \hat H = - \hat g^{-{1\over 4}} ( i \hat p_\m + \hat A_\m )
\hat g^{1\over 2} \hat g^{\m\n} ( i \hat p_\n + \hat A_\n )
\hat g^{-{1\over 4}} + \hat S.}
{}From (3.4) it follows that the trace in \ty\ can be rewritten as
\eqn\newty{ I = \lim_{\b \to 0}
\sum_{j} \int d x {\sqrt g}\
< \! x,j\vert \e^{- \beta  \hat H} \vert x,j\! > }
The  gauge transformation properties in \trprop\
 for the transition amplitude as well
as its scalar character under reparametrization will
be the defining principles for our construction of the path integral
representation, to which we now turn.

We are going to define the  path integral representation
for the transition amplitude,
to be considered as a matrix with indices $j_f$ and $j_i$,
for a particle starting
at point $q^\m_i$ at the initial time $t_i$ and reaching point $q^\m_f$
at the final time $t_f$:
\eqn\trans{ \langle q^\m_f,j_f,t_f | q^\m_i,j_i,t_i \rangle
= < \! q^\m_f,j_f\vert \e^{- (t_f-t_i)  \hat H} \vert q^\m_i,j_i
\! >  .}
 The path integral  is defined as follows
\eqn\cpi{\eqalign{
& \langle q^\m_f,j_f,t_f | q^\m_i,j_i,t_i \rangle   =
\int_{q(t_i) = q_i}^{q(t_f)=q_f} (\tilde {\cal D} q)
\ T\ \e^{ - S[q] } \cr
&S[q] = \int_{t_i}^{t_f} dt\ \biggl (
{1\over 2} g_{\m\n} \dot q^\m \dot q^\n + A_\m \dot q^\m + V \biggr )\cr
& (\tilde {\cal D} q)
= \prod_{t_i< \! t < \! t_f} {\sqrt{ g(q(t))}} d q(t),\cr}}
where $T$ denotes time-ordering along the histories  $q^\m(t)$.
The vector potential $A_\m$ and the scalar
potential $V$ are matrices,
so that the action itself is a matrix (hence the explicit indices
$j_i$ and $j_f$
shown explicitly on the left-hand side of the first equation,
but omitted in the
right-hand side for notational convenience).
The explicit relation between
the scalar potential
$V$ in \owe\ and \cpi\ and the scalar potential $S$  appearing in \ty\
will be given later.
Note that the notation $\langle \  |... | \ \rangle$ means
averaging with the functional integral under the boundary
conditions given in the brackets,
while $< \! |...| \! >$
will denote the usual bras and kets of operatorial
quantum mechanics, so that \trans\ is the statement
of the equivalence of the operatorial formalism
and the functional integral formalism.
To check the correctness of this
 definition  of our path integral,
we must check its  transformation properties under
 reparametrizations and gauge variations.
As far as reparametrizations are concerned, we can immediately see
that the path integral is covariant (in fact it is a scalar),
since it is built from
manifestly covariant  objects.
Turning to the gauge symmetry, we can compute the variation of
 the transition amplitude
 under an
infinitesimal gauge variation ($U= \e^\Lambda$ with
$\Lambda$ infinitesimal)
\eqn\check{\eqalign{
\delta \langle q^\m_f,j_f,t_f | q^\m_i,j_i,t_i \rangle &=
 \langle q^\m_f,j_f,t_f |\bigl ( \Lambda (q_f) - \Lambda (q_i)
\bigr ) \vert
 q^\m_i,j_i,t_i \rangle\cr
&= \Lambda_{j_f}{}^k (q_f)
 \langle q^\m_f,k,t_f | q^\m_i,j_i,t_i \rangle -
 \langle q^\m_f,j_f,t_f
| q^\m_i,l,t_i \rangle \Lambda_l{}^{j_i} (q_i).
\cr}}
The first line is a consequence
of the transformation properties of  time-ordered
exponentials.
For the finite case it reads
\eqn\finite{  \langle q^\m_f,j_f,t_f | q^\m_i,j_i,t_i \rangle' =
U_{j_f}{}^k (q_f)
 \langle q^\m_f,k,t_f | q^\m_i,l,t_i  \rangle (U^{-1})_l{}^{j_i} (q_i), }
which is the correct transformation property expected from the
operatorial picture.

We now turn to the path integral measure of \cpi.
A key point introduced in \fb\ is to exponentiate
the non-trivial part of the measure
using ghost fields, thereby leaving  a
translational-invariant measure
necessary for performing the perturbative expansion.
We present it here in a modified form, which
helps in keeping track of the cancellation of  potentially divergent
terms.
The latter are
 typically present in non-linear models of quantum mechanics
(see ref. \lee).
Since we need to exponentiate the factor $\sqrt g$
present in the measure of \cpi, we introduce anticommuting ghost
fields $b^\m,c^\m$  and commuting ghost fields $a^\m$
coupled to the metric $g_{\m\n}$.
The effect of path integrating over $b^\m,c^\m$ will be to produce
a factor $g$, while the path integration over $a^\m$ will
reduce it to $\sqrt g$, thus recovering the correct
measure.
Therefore the  covariant path integral
which we are going to compute
perturbatively in the next section
reads as follows
\eqn\ncpi{\eqalign{&\langle q^\m_f,t_f | q^\m_i,t_i \rangle =
\int_{q(t_i) = q_i}^{q(t_f)=q_f}
({\cal D}q)({\cal D}b)({\cal D}c)({\cal D}a)
\  T \e^{ - S[q,b,c,a] } \cr
&S[q,b,c,a] = \int_{t_i}^{t_f} dt\ \biggl (
{1\over 2} g_{\m\n} (\dot q^\m \dot q^\n + b^\m c^\n + a^\m a^\n)
+ A_\m \dot q^\m + V \biggr )
\cr
&({\cal D}q) = \prod_{t_i< \! t < \! t_f} d q(t),\ \
{\rm  similarly\ for\ }
({\cal D}b)  \
({\cal D}c) \
({\cal D}a). \cr}}

\newsec{Perturbation expansion}

We now present
the perturbative evaluation of the covariant path
integral defined in the previous section.
Our strategy will be to make full use of the general
coordinate invariance and pick the
coordinate system best suited for carrying
out the  computation, namely the Riemann
normal coordinates centered at the
final point $q^\m_f$.
Before proceeding further, we introduce the dimensionless time variable
$\tau = {t -t_f \over \beta}$, with $\beta=t_f-t_i$, and rescale
the ghost fields as follows $ (b^\m, c^\m, a^\m) \to {1\over\beta}
 (b^\m, c^\m,a^\m)$,
so that $\beta$ will turn into a loop-counting parameter,
nicely organizing  the loop expansion.
We do not keep track of a possible Jacobian
for this rescaling, since the overall
normalization will later be determined by other means.
After such redefinitions
the path integral takes the following form
\eqn\nncpi{\langle q^\m_f,j_f,t_f | q^\m_i,j_i,t_i \rangle =
\int_{q(t_i) = q_i}^{q(t_f)=q_f}
({\cal D}q)({\cal D}b)({\cal D}c)({\cal D}a)
\ T \e^{ - {1\over \beta} S[q,b,c,a] },}
where
\eqn\pvns{S[q,b,c,a]=\int_{-1}^{0}d\t \ \biggl [{1 \over 2}
g_{\m\n}(q)(\dot q^{\m}\dot q^{\n}
+b^{\m}c^{\n}+a^{\m}a^{\n})+\b A_{\m}(q) \dot q^{\m} +\b^{2}V(q)
\biggr ].}

We now go over to Riemann normal coordinates by introducing functions
$q^{\m}(\t,\l)$ which are geodesics in $\l$ for fixed $\t$, and which
start at $q^{\m}_{f}$ and end at $q^{\m}(\t)$:
\eqn\pvnqm{q^{\m}(\t,\l=0)=q^{\m}_{f} , \ \ \ \
\ q^{\m}(\t,\l=1)=q^{\m}(\t).}
The normal coordinates $z^{\m}(\t)$ for a point with coordinates $q^{\m}(\t)$
are then by definition
\eqn\pvnzm{z^{\m}(\t)={\pa \over \pa \l}q^{\m}(\t,\l) \biggm|_{\l=0}.}
The action $S[q(\t,\l=1)]$ can be Taylor expanded in $\l$ around $\l=0$,
and if the action is an Einstein scalar, one can replace ordinary
$\l$ derivatives by covariant derivatives ${D \over {\pa\l}} ={\pa q^{\m}
\over
{\pa \l}} \nabla_{\m}$:
\eqn\pvnssum{S=\sum_{n=0}^{\infty}{1 \over n!}
\biggl ({D \over {\pa \l}} \biggr )^{n}S[q(\t,\l)]
\bigg|_{\l=0}.}
Since the $q(\t,\l)$ are geodesics, $({D \over {\pa\l}})^{2}q(\t,\l)=0$, and
in the expansion either no derivatives act on $q^{\m}(\t,\l)$ (yielding the
$\t$-independent constant $q^{\m}_{f}$) or one derivative (yielding
$z^{\m}(\t)$).
Another useful identity is
\eqn\pvni{{D \over {\pa\l}}{D \over {\pa\t}}z^{\m}(\t,\lambda)=
\biggl [{D \over {\pa\l}},{D \over {\pa\t}} \biggr ]
z^{\m}(\t,\lambda)
=R_{\a\b}{}^{\m}{}_{\n}z^{\a}\dot q^{\b}z^{\n},}
where ${D \over {\pa\t}}={\pa q^{\m} \over {\pa\t}}\nabla_{\m}$ and
$z^\m (\t,\lambda)= {\pa \over {\pa \lambda}} q^\m (\t,\lambda)$.
Of course, ${D\over {\pa \lambda}} \dot q^\m (\t,\lambda) = \dot z^\m
 + O(\lambda)$.

Using this expansion of the action into normal coordinates,
%and shifting
%the origin of the q-coordinate system to $q^{\m}_f$,
one obtains
\eqn\expan{\eqalign{ S = \int_{-1}^0 d\tau &\biggl [
{1\over 2} \biggl ( g_{\m\n}(0) + {1\over 3} R_{\a\m\n\b}(0) z^\a z^\b
+{1\over 6} \nabla_\g R_{\a\m\n\b}(0) z^\g z^\a z^\b \cr
&+ R_{\g\delta\a\m\n\b}(0)
z^\g z^\delta z^\a z^\b + \cdots  \biggr ) (\dot z^\m \dot z^\n
+b^\m c^\n + a^\m a^\n )\cr
&+\b \biggl (  A_\m (0)  + \nabla_\n A_\m (0) z^\n
+ A_{\l\n\m} (0) z^\l z^\n + A_{\a\b\g\m}(0)
z^\a z^\b z^\g  + \cdots \biggr )
\dot z^\m \cr
&+ \b^2 \biggl (
V(0) + z^\m \nabla_\m V(0) + {1\over 2} z^\m z^\n \nabla_\m
\nabla_\n V(0) + \cdots \biggr ) \biggr ] , \cr}}
where the following tensors have been defined
\eqn\tensor{\eqalign{ & R_{\g\delta\a\m\n\b} =
{1\over{20}} \nabla_\g \nabla_\delta R_{\a\m\n\b} +
 {2\over{45}} R_{\a\m\s\b} R_{\g\n}{}^\s{}_\delta \cr
& A_{\l\n\m} =
{1\over 2}  \nabla_\l \nabla_\n A_\m  + {1\over 6}
 R_{\l\m}{}^\s{}_\n A_\s \cr
&A_{\a\b\g\m}=  {1\over 6}  \nabla_\a \nabla_\b \nabla_\g
  A_\m + {1\over 6} R_{\b\m}{}^\r{}_\g \nabla_\a A_\r +
{1\over 12} A_\r \nabla_\a R_{\b\m}{}^\r{}_\g . \cr}}
Note the $\beta$ dependence of the vector and scalar potentials produced
by the introduction of the time variable $\t$.
It implies that they will start contributing
at higher loops with respect to the ``interactions''
originating from the metric tensor.
At this stage, we could start the perturbative quantum computation
by splitting the field $z^\m$ into a classical
and  a quantum  piece
\eqn\split{ z^\m = z^\m_{cl} + z^\m_{qu} .}
The classical path $z^\m_{cl}$ is taken to be
a solution of the classical
equation of motion in the limit $\beta \to 0$, which is the
geodesic equation.
So $ z_{cl}^\m (\t) = - z_i^\m \t$, where
$z_i^\m$ are the Riemann normal coordinates
of the initial point $q_i^\m$.
It takes into account  the boundary conditions
prescribed to the path integral. The quantum fluctuations
$z^\m_{qu}$ therefore vanish at $\tau = -1,0$
and can be  Fourier-expanded  as follows
\eqn\sinuses{z^\m_{qu} = \sum_{n=1}^{\infty} z^\m_n \sin
(\pi  n \tau).}
A similar expansion holds also for the ghost fields, which are
zero at the boundary
\eqn\ghostexp{
b^\m = \sum_{n=1}^{\infty} b^\m_n \sin (\pi  n \tau) ;\ \ \
c^\m = \sum_{n=1}^{\infty} c^\m_n \sin (\pi  n \tau) ; \ \ \
a^\m = \sum_{n=1}^{\infty} a^\m_n \sin (\pi  n \tau) .}
As we explained,
these boundary conditions follow
 from the fact that  the measure factor $\sqrt  g$ is not included in
\cpi\ for $t_i$ and $t_f$.
 The  evaluation of the transition amplitude in the two-loop
approximation is enough to determine the quantum
Hamiltonian associated with
the path integral. Put differently,
a two-loop computation is sufficient to verify that the transition amplitude
satisfies a Schr\"odinger equation
(diffusion equation)
\eqn\Sc{\eqalign{ & {d \over {dt}}
 \langle x^\m, i, t| y^\m,j,t' \rangle = - H
\langle x^\m, i, t| y^\m,j,t' \rangle
 \cr & H =  -{1 \over 2}(\nabla^\m + A^\m)(\partial_\m + A_\m)
- {1 \over 8}R
+ V .\cr }}
where in the first equation $H$ acts on the indices $x^\m $ and $i$.
Such a computation was presented in \fb\ and will not be reproduced here,
 since
the introduction of the time ordering does not bring in any
modification at this stage. The above Hamiltonian shows that the
connection with the operatorial formulation presented in sect. 3
is achieved by setting  the scalar potential
$ V = {1\over 2} S + {1\over 8 } R$.
An additional outcome of the above procedure is the determination of
the complete normalization of the path integral measure
\eqn\pim{\eqalign{ {\cal D}q {\cal D}b {\cal D}c {\cal D}a &=  A
\prod_{m=1}^{\infty}
\prod_{\m=1}^d ( \pi m^2  )^{1\over 2}
d z_m^\m  d b^\m_m d c^\m_m d a^\m_m  \cr
A &= {1\over { (2 \pi \b)^{{d\over 2}}}} \cr}}
A consequence of this result is that all Gaussian integrals over
the modes \underbar{and} the final integration over $q_{i}^{\m}$ in \opvn\
 are normalized to unity if one uses in \expan\ only the terms with
$g_{\m\n}(0)$ and puts in \opvn\ the $q_i$ in $\sqrt {g(q_i)}$  and
$\psi(q_i,t_i)$ equal to $q_f$.
(The kinetic terms in the action are read off from \expan\
and are given by
\eqn\kin{ S_{_{(2)}} ={1\over 2} g_{\m\n} (0)  z^\m_i z^\n_i
+ {1\over 4} g_{\m\n} (0)
 \sum_{n=1}^\infty (\pi^2 n^2 z^\m_n z^\n_n
+ b^\m_n c^\n_n + a^\m_n a^\n_n ).}
The factors $\sqrt g$ due to integration over
$z_m^\m,b_m^\m,c_m^\m$ and $a_m^\m$ cancel, whereas the integration
over $z_i^\m$ produces a factor $(2\pi \beta)^{{d \over 2}}
(g)^{-{1 \over 2}}$, which
cancels  against the
 $(g)^{1 \over 2}$ in \opvn\ and $A$ in \pim.)

Having established  the equivalence \trans\ for
$ V ={1\over 2} S + {1\over 8 } R $, we proceed to the task
of computing the general trace in \ty\
using the path integral representation
\eqn\pir{I = \lim_{\b \to 0}
{\rm Tr}\ \e^{-\b H} = \lim_{\b \to 0} \sum_i \int dx {\sqrt g}\
\langle x^\m,i,0 | x^\m,i,-\beta \rangle.}
The evaluation of the right-hand side
proceeds as follows.
The classical path in \split\ is now $ z^\m_{cl} = 0$,
and we can drop the subscript in $ z^\m_{qu}$ without
possibility of confusion.
{}From the quadratic piece in \kin\ we recognize the following
propagators
\eqn\propag{\eqalign{
&\langle z^\m(\tau) z^\n(\tau') \rangle
= - \beta  g^{\m\n}(0) \Delta (\tau,\tau') \cr
&\langle b^\m(\tau) c^\n (\tau')\rangle
=  - \beta  g^{\m\n}(0) \bigl ( 2\pa_\tau^2 \Delta (\tau,\tau')
\bigr )\cr
&\langle a^\m(\tau) a^\n(\tau') \rangle
= - \beta  g^{\m\n}(0) \bigl (- \pa_\tau^2 \Delta (\tau,\tau')
\bigr ), \cr}}
where
\eqn\del{\Delta (\tau,\tau') = \sum_{n=1}^{\infty} \biggl [ -
{2 \over {\pi^2 n^2}} \sin (\pi n \tau) \sin (\pi n \tau') \biggr ].}
Note that we can sum the series in \del\ and obtain
\eqn\func{\Delta (\tau,\tau') = \tau ( \tau' +1) \Theta (\tau - \tau')
+ \tau' ( \tau +1) \Theta (\tau' - \tau), }
where the theta function is defined as usual by
\eqn\th{\eqalign{  & 0 \ \ \ {\rm for }\ \tau < \  \tau' \cr
\Theta (\tau - \tau') =
\  & {1\over 2} \ \ \ {\rm for }\ \tau = \tau' \cr
 & 1 \ \ \ {\rm for }\ \tau \ > \tau' \cr}}
Indeed, ${d^2 \over d\t^2}\Delta(\t,\t')=\d(\t-\t')$,
 while $\Delta(0,\t')=
\Delta(\t,0)=0$. Note that $\Delta(\t,\t')$ is continuous
at $\t' \downarrow \t$
and $\t' \uparrow \t$, and $\Delta(\t,\t)=\t(\t+1)$. Also
\eqn\pvndott{
{d \over {d\t}}\Delta(\t,\t') \equiv{}^{\bullet}\!\Delta(\t,\t')
=\t'+\theta(\t-\t')}
\eqn\pvnddot{
{d \over {d\t'}}\Delta(\t,\t') \equiv \Delta^{\bullet}(\t,\t')
=\t+\theta(\t'-\t)}
and also $\Delta^{\bullet}
={}^{\bullet}\!\Delta={1 \over 2}+\t$ at $\t'=\t$.
However, we must recall that the path integral is naturally regulated
by truncating the mode expansion at a fixed mode $M$, so that
the derivatives in \propag\ should be taken before the limit
$M \to \infty$. At fixed  $M$, all propagators are well-defined
functions, with finite limits at coinciding points
$\tau = \tau'$.

The interaction vertices are  given by the
remaining pieces of  the action, $S_{_{int}} = S - S_{_{(2)}}$,
with $S$ given in \expan\ and $ S_{_{(2)}}$  in \kin.
The final result for the basic trace is thus as
follows. In $d$-dimensional Euclidean space,
\eqn\kak{ I^{(d)} = \lim_{\b \to 0}  {\rm Tr }\  {\rm e}^{-\beta H}
= \lim_{\b \to 0 } {1\over {(2 \pi \b)^{d\over2}}}  \int dx {\sqrt g(x)}
\langle 0 | T \e^{-{1\over \b} S_{_{int}}} | 0  \rangle ,}
where
\eqn\kbk{\eqalign{ S_{_{int}}=&
\int_{-1}^0 d\tau \biggl [
{1\over 2} \biggl (  {1\over 3} R_{\a\m\n\b}(x) z^\a z^\b
+ \dots  \biggr ) (\dot z^\m \dot z^\n
+b^\m c^\n + a^\m a^\n )\cr
&+\b \biggl (  A_\m (x)  + \nabla_\n A_\m (x) z^\n
+  \cdots \biggr )
\dot z^\m
+ \b^2 \biggl (
V(x) + z^\m \nabla_\m V(x)  + \cdots \biggr ) \biggr ]. \cr}}
One must evaluate in $d$ dimensions only all \lq\lq vacuum
polarization\rq\rq
graphs that are proportional
to $\b^{d\over 2}$.  They are obtained by expanding
$ \e^{ ( -\b^{-1} S_{_{int}})} $,
and contracting all  $ z^\m, b^\m, c^\m$ and $ a^\m$,
using the propagators in \propag.
The  trace anomaly is then obtained by multiplying
the result by the Weyl weights of the fields.

\newsec{ Computation of the trace anomalies in $d=2$ and $d=4$}

To compute the trace anomalies in $d=2$, $ I^{(2)}$ in \kak,
we need
 the $\b$ term in \kak. Since propagators go like $\b$
and vertices like ${1 \over \b}$ (or $\b^0$ for $A$, or  $\b$ for $V$),
 this means that we must compute 2-loop diagrams in $d=2$.
Similarly
for  $d=2k$  we need the $k+1$ loop
contribution.
 Using the above propagators and vertices,
we get for $d=2$ only a contribution
from the ${1 \over 3}R_{\a\m\n\b}$ vertex in \expan.
The corresponding diagrams have the topology of the number 8,
with none or one of the loops containing a ghost.
The result is
proportional to
\eqn\pvnddt{ \int^0_{-1}  d\t \biggl [ ({}^{\bullet}\!\Delta^{\bullet}
+\Delta^{\bullet \bullet})\Delta
-\Delta^{\bullet} \Delta^{\bullet} \biggr ]_{\t=\t'}. }
The last term is clearly finite.
Partially integrating  the first term by
using the fact  that
$ ({}^{\bullet}\!\Delta^{\bullet}
+\Delta^{\bullet \bullet}) = \pa_\t
({}^{\bullet}\!\Delta  \ {\rm at} \ \t =\t')$
immediately shows that \pvnddt\
is finite. Also in higher loops finiteness is manifest thanks to using
vector ghosts. The result is
\eqn\tratwo{ I_{(2)} = {\rm tr}_{_{YM}}
\int {d^2 x \over {2 \pi }} {\sqrt g}\
\biggl [  {1\over{24}} R - V
\biggr ],}
where the trace is over the finite dimensional representation space
of the gauge group.

The computation for obtaining the  $d=4$ result
is more involved, but nevertheless straightforward.
Let us first quote the result and then briefly comment on its derivation.
\eqn\trafour{ \eqalign{
I_{(4)} =  {\rm tr}_{_{YM}}
\int {d^4 x \over {(2 \pi )^2}} {\sqrt g}\
&\biggl [ {1\over {720}} \bigl ( R_{\a\m\n\b}R^{\a\m\n\b} - R_{\m\n} R^{\m\n}
\bigr )  + {1\over{480}} \nabla^2 R
 -{1\over 12} \nabla_A^2 V
\cr
&+ {1\over 2 }
\biggl ( {1\over{24}} R - V \biggr )^2 + {1\over{48}} F_{\m\n}F^{\m\n}
\biggr ]. \cr}}
The $A_\m$ contribution due to one vertex $-{1 \over \b}S_{_{int}} $ vanishes,
since $ \int_{-1}^0 d\t^{\ \bullet}\!\Delta(\t,\t) $ as well as
$\int_{-1}^0 d\t  \Delta(\t,\t){}^{\bullet}\!\Delta(\t,\t) $
 are zero.
The $A_\m$ contribution coming from two vertices $-{1 \over \b}S_{_{int}}$
produce the kinetic terms in ${1 \over 48}F^2_{\m\n}$. This comes
about because $\int \! \int d\t d\t'\langle
(z^\n\dot z^\m)(\t)\ (z^\s\dot z^\r)(\t')
 \rangle $ is proportional to $\int\!\int d\t d\t'
 [g^{\n\s}g^{\m\r}
\Delta(\t,\t'){}^{\bullet}\!\Delta^{\bullet}(\t,\t')+g^{\n\r}g^{\m\s}
 \Delta^{\bullet}(\t,\t'){}^{\bullet}\!\Delta(\t,\t')  ]$,
and partially integrating $\ ^{\bullet}\!\Delta^{\bullet}$, using that $\Delta$
vanishes at the boundary, gives the curl structure of the Maxwell action. The
$(DA)AA$ terms in the Yang-Mills action are a good test that our time-ordering
prescription is required by gauge invariance. They are treated like the $AV$
and $AAV$ terms, see below.
The $V$ terms in \trafour\ come from the vertex
$- \b [V +{1 \over 2}z^\m z^\n \nabla_\m \nabla_\n V]$.
First there is the term
${1 \over {(2\pi \b)^2}}{1 \over 2}[\b V]^2={1 \over {8\pi^2}}V^2$.
The term with $\nabla^2 V$ comes of course from
$\langle z^\m z^\n \rangle \nabla_\m \nabla_\n V$.
It is instructive to see how
the covariantizations in $-{1 \over {12}}\nabla^2_A V=-{1 \over {12}}\pa^\m
(\pa_\m V+[A_\m,V]) -{1 \over {12}}\biggl [ A^\m,\pa_\m V+[A_\m,V] \biggr ] $
arise, because here we see the crucial role of our time-ordering prescription.
For example, from
\eqn\lpvn{\eqalign{&T(-{1 \over \b})^2{1 \over {2!}}
 2\int_{-1}^0 \b(A_\m \dot z^\m
+\nabla_\n A_\m z^\n \dot z^\m)d\t \int_{-1}^0 \b^2(V+z^\r \pa_\r V)d\t' \cr
& =\b A_\m\pa_\r V\int_{-1}^0 \dot z^{\m}d\t \int_{-1}^\t z^\r d\t' +
\b \pa_\r V A_\m \int_{-1}^0 z^\r d\t \int_{-1}^\t \dot z^\m d\t'. \cr}}
Partially integrating and using $z^\m(0)=z^\m(-1)=0$
yields indeed $-{{\b^2} \over 6}[A^\m,\pa_\m V]$. Similarly the
$[\pa^\m A_\m,V]$ is recovered. The most interesting term is
$\biggl [A_\m,[A_\m,V] \biggr ]$; one now has three ways of ordering the
time coordinates $\t,\t',\t''$, and one indeed obtains the combination
$A_\m A^\m V-2A_\m V A^\m + V A_\m A^\m$, and with the correct coefficient.

We have been explicit in these calculations, because they show the
importance of time ordering in order to obtain gauge-invariant results.

Having computed the general traces \tratwo\ and \trafour,
we can specialize them to deduce the Weyl anomalies
of the various fields discussed in sect. 2.
We need to use $V={1 \over 8}R-{1 \over 2}\xi R$
for the spin 0 regulator, and
$V=-{1 \over 4}F_{\m\n}\g^{\m\n}$ and $A_{\m} \rightarrow A_\m+\omega _\m$
for the spin ${1 \over 2}$ regulator, obtaining the following result
for the trace anomalies
of a real scalar and a complex spin ${1 \over 2}$ field in two dimensions:
\eqn\ftra{ A_0^{(2)} =
A_{1\over 2}^{(2)} = - \hbar {\rm tr_{_{YM}}} \int
{d^2 x \over {2 \pi }} {\sqrt g}
{1\over {12}} R
 \s. }
In four dimensions we have instead
\eqn\ftraf{ A^{(4)} = \hbar
{\rm tr_{_{YM}}}\int {d^4 x \over {(2 \pi )^2}} {\sqrt g}
\biggl [
 a  R_{\a\m\n\b}R^{\a\m\n\b}
+b  R_{\m\n} R^{\m\n}
+c R^2
+d \nabla^2 R
+e F_{\m\n}F^{\m\n}
\biggr ] \s}
with the various fields contributing as follows
\eqn\last{\vcenter{\halign{#\hfil&#\hfil&#\hfil&#\hfil&#\hfil&#\hfil\cr
spin 0 & $a={1\over{720}}$ &
$b=-{1\over{720}}\ $ & $c=0$ & $d=-{1\over{720}}\ $
  & $e={1\over{48}}$ \cr
spin ${1\over 2}$ (Dirac spinor) \ & $a={7\over {1440}}\ $
  & $b={1\over{180}}\ $
  & $c= -{1\over {288}}\  $ & $d= -{1\over {120}} $ & $e= {1\over {6}}$ \cr
spin 1 (Abelian) & $a=-{13\over{720}}$
  & $b={11\over {90}} $ & $c=-{5\over {144}}$ & $d={1\over {40}} $
  & $e= 0 $ \cr
spin 1 (non-Abelian) &  $a=-{13\over{720}}\ $
  & $b={11\over {90}}$ & $c=-{5\over {144}}$
  & $d={1\over {40}}$ & $e= -{11 \over {24}}$ \cr}}}
where in addition we have used for the
ghost regulator  $V= {1\over 8} R$ and for the
gauge regulator
 $V={1 \over 8}R\d_m{}^n-{1 \over 2}R_m{}^n
-{1 \over 2 }F_m{}^n$
and  $A_{\m} \rightarrow A_\m+\omega_\m$,
with $\omega_\m$ acting in the vector representation.
These results agree with the literature \duff.
For completeness, we recall that eq. \ftraf\ can also be written in the
following form
\eqn\newform{
 A^{(4)} = \hbar
 \int {d^4 x \over {(2 \pi )^2}} {\sqrt g}
\biggl [
\a {\cal F}
+\b {\cal G}
+\g \nabla^2 R
+e ({\rm tr}_{_{YM}} F_{\m\n}F^{\m\n})
\biggr ] \s,}
where  ${\cal F}= R_{\a\m\n\b} R^{\a\m\n\b} -2 R_{\m\n} R^{\m\n} +{1\over 3}
R^2$ is the square of the Weyl conformal tensor and ${\cal G} =
 R_{\a\m\n\b} R^{\a\m\n\b} -4 R_{\m\n} R^{\m\n} + R^2$ is the topological
Euler density. The coefficients $\a, \b, \g$
are given by
\eqn\newcoef{
\eqalign{\a &= {1\over{480}} (N_0 + 6 N_{1\over 2} + 12 N_1)\cr
\b &= -{1\over {1440}} (N_0 + 11N_{1\over 2} + 62 N_1)\cr
\g &= -{1\over {720}} (N_0 + 6N_{1\over 2} -18 N_1), \cr}}
where $N_0, N_{1\over 2}$ and $N_1$
are the number  of real spin $0$, complex
(Dirac) spin ${1\over2}$, and real spin $1$ fields, respectively.
This form of presenting the trace anomaly clearly
shows that the coefficients in eq. \ftraf\ are not all independent \duff.
This is a consequence of the Wess-Zumino consistency conditions,
which guarantee
the  absence of a single  $R^2$ term in the anomaly
\ref\bon{L. Bonora,
P. Cotta-Ramusino and C. Reina,
 Phys. Lett. B126 (1983) 305.}.
Moreover, the  coefficient  $\g$ in \newform\ can be modified at will
(in  particular can be set to zero)
by adding to the effective action
a finite local counterterm proportional
to $R^2$. For spin 1, only the sum of the contributions of the non-ghost
sector and the ghost sector satisfies the consistency conditions.
The gauge-fixed action is not Weyl-invariant, but its Weyl variation
is BRST exact and
has vanishing expectation value \nil.

\newsec{Conclusions}

In this article we have computed trace anomalies in quantum field theory
by using path integrals in quantum mechanics. Our work extends
the analysis of Alvarez-Gaum\'e and Witten of chiral anomalies, but
because trace  anomalies are not  of a topological nature,
certain fundamental and interesting aspects of path integrals
had to be taken into account, which could be neglected
in the case of chiral anomalies. As already indicated in
\ref\new{K. Fujikawa, S. Ojima and S. Yajima,
Phys. Rev. D34 (1986) 3223.}, this makes  the evaluation
of trace anomalies more complicated than that of chiral anomalies.
Of interest is the method itself, rather than the results.
The quantum mechanical approach is an excellent tool
for computing anomalies. It avoids the painful Baker-Campbell-Hausdorf
expansion of  a direct application of  Fujikawa's method.
It is clearly related  to the DeWitt's heat-kernel approach \ref\vil
{G.A. Vilkovisky, preprint CERN-TH.6392/92.},
but it avoids the  bilocal tensor calculus of the latter.

We have given  a general formula, eq. \kak, from which trace
anomalies in any dimension can be obtained. We then computed the
trace anomalies in $d=2$ and $d=4$ for spin $0, {1\over 2}$ and $1$
at one loop with external gravitational and Yang-Mills fields.
Our results agree, of course, with the literature \duff.
In the literature also the trace anomalies for $d=6$ and $d=8$
have been obtained (by Gilkey \ref\gilk{ P.B. Gilkey,
J. Diff. Geom. 10 (1975) 601\semi
For the case $d=4$
see Proc. of Symposia in Pure Math. 27 (1975) 265 (published by
Amer. Math. Soc.).}
and Avramidi \ref\avra{ I.G. Avramidi, Theor. Math. Phys. 79
(1989) 219.}, respectively, who both used  the heat-kernel approach).
It would be interesting to find an explicit closed formula
for  the trace anomalies in any $d$, similar to the result of \ag\
based on the \lq\lq Dirac genus\rq\rq \
for the chiral anomalies.
Perhaps our eq. \kak\ could be a starting point.

In the case of the chiral anomaly, one needs  in $d$ dimensions
$d$ powers of the  fermionic zero modes $\psi^\m_0$
in  order  that the Berezin integration is non-vanishing.
This introduces a factor $\b^{d\over 2}$, which cancels
the factor $\b^{-{d\over 2}}$ in eq. \pim.
For the trace anomaly, no $\psi^\m_0$ are present, and hence one must  make a
loop expansion
on the world line.

In the transition from phase space to configuration space path integrals,
a measure $\sqrt g$ is produced, familiar from non-linear sigma
models.
We have treated this measure in a way that is different from the usual
$\delta(0)$ approach \lee. Namely, rather than add extra vertices to the
action, which are proportional to ln det $g$
times $\delta(0)$ and which are supposed to cancel
leading  divergences, we exponentiated this $\sqrt g$
to a perfectly
regular term in  the action with new ghost fields \fb,
similar to the familiar Faddeev-Popov ghost action.
It was then clear, diagram by diagram, how all the  divergences
(not only the leading ones)
cancel in all loops for finite $\b$.
An example of such a cancellation was given in \pvnddt.
Of course, we need  to exponentiate the factor $\sqrt g$  in order to be left
with
 a translationally invariant measure from which  propagators can be
defined.
Historically, these factors $\delta(0)$ first occurred in the proof
of Matthews's theorem
\ref\mat{P.T. Matthews, Phys. Rev. 76 (1949) 684\semi
T. Duncan and C. Bernard, Phys. Rev. D11 (1975) 848.},
 which states  that the Hamiltonian
and the Lagrangian canonical quantization are equivalent if one adds in the
latter extra vertices proportional to $\delta(0)$, and in the  analysis
of Lee and Yang
\ref\LY{T.D. Lee and C.N. Yang, Phys. Rev. 128 (1962) 885.}
 of massive vector bosons. We believe that
 our approach is simpler and clearer.

In the canonical formalism, Heisenberg fields satisfy the field equations,
and fields must be expanded  in a complete set of  solutions.
In the path integral, fields  do not need to satisfy any field equations,
and one is free to expand  in any basis. We found it convenient
to decompose  the coordinates $q^\m(t)$ into a background piece,
which satisfies the  field equation  (a geodesic),
and fluctuations. We expanded  all fluctuations in the trigonometric
basis $\sin (\pi m \tau)$ with $m=1,2,3,\dots$.
This raises the question whether  our results depend on the choice of basis.
Since an ambiguity is known to exist if one uses time discretization,
and is fixed by imposing gauge invariance, we expect our results
to be basis-independent as long as  they are Einstein- and gauge-invariant.

We end with a comment on Feynman diagrams,  which provide
the oldest method of computing anomalies and which have been
used with success for chiral anomalies \ag.
The simplest diagram  which one might consider to yield a trace
anomaly is the two-point function in $d$ dimensions,
with candidate trace anomaly $ (\nabla^2 )^{{d\over 2} - 1} R$.
In  $d=2$ it is a genuine anomaly, but for  $2d\! >2$
it can be removed by a counterterm with two factors $R$
(as observed by Adam Schwimmer (private communication)
the most general two-point function
which is both transversal and traceless is proportional to
$ {1\over 2} (d-1) ( \theta_{\m\r} \theta_{\n\s} +
\theta_{\m\s} \theta_{\n\r}) - \theta_{\m\n} \theta_{\r\s}$,
where $ \theta_{\m\n} = \eta_{\m\n} - k_\m k_\n k^{-2}$,
but in $d=2$ this expression identically vanishes).
The triangle diagram cannot reproduce
all $h^2$ terms in $d=4$, because the $h^2$ terms
of the Euler invariant vanish in all dimensions.
Thus, in $d=4$, one would need  to compute the
4-point function, which is tedious to evaluate.
Moreover, in higher $d$ there are more invariants, which
constitute the trace anomaly, and hence further  diagrams
would have to be computed. (For the chiral anomaly,
only one diagram  needs to be calculated  in a given dimension.)
We conclude  that the  quantum mechanical approach is a much simpler
way to compute the trace anomalies.

\vskip 1cm

{\bf Acknowledgements}

F.B. thanks CERN for hospitality and the
``Blanceflor Boncompagni-Ludovisi Foundation''
for supporting his stay at CERN.
\vfill\eject

\appendix{A}
\indent
In the computation of loops on the world line,
we have used the explicit form of
the propagator function $ \Delta(\t,\t') $.
As a result, all loop integrals
became trivial, typically powers of $\t$ and $\t'$ from $-1$ to $0$.
However, if one is interested in truncated results one needs to use the
representation of  $ \Delta(\t,\t') $
in terms of sine functions, and then the loop integrals
for $d=4$ become (when the
truncation is removed at the end of the
calculation) proportional to the following
double sums
\eqn\A{\sum_{n,m =1}^{\infty} {1\over { (m+n)^2 mn}}={\pi^4 \over {180}};\
\sum_{n,m =1}^{\infty} {1\over { (m+n) m^2 n}}={\pi^4 \over {72}};\
\sum_{n,m =1}^{\infty} {1\over { (m+n)^2 m^2}} = {\pi^4 \over {120}}.}
The sums $\sum_{n=1}^{\infty} n^{-p}$
are given by the polylogarithms ${\rm Li}_{2p-2} (1)$;
for example ${\rm Li}_2 (1) = {\pi^2 \over 6}$,
${\rm Li}_4 (1) = {\pi^4 \over {90}}$, ${\rm Li}_6 (1) = {\pi^6 \over {945}}$
and ${\rm Li}_8 (1) = {\pi^8 \over {9450}}$, see \ref\Levin{L. Levin,
\lq\lq Polylogarithms and associated functions\rq\rq
 (Elsevier, North-Holland,
1981).}.
Since $\sum_{n=1}^{\infty} n^{-2} ={\pi^2 \over 6}$, the factor
$\pi^4$ in \A\ is plausible, and one easily finds the coefficients
${1\over {180}}, {1\over {72}}$ and ${1\over {120}}$
 from a numerical
approximation. However, since (to our surprise) the results
 for these double sums
are not found in the  standard
references \ref\M{M. Abramowitz and I.A. Stegun,
\lq\lq Handbook of mathematical functions\rq\rq, 1968.},
we give an analytical derivation, which can also be generalized
to more complicated sums.

First, consider $I(p,q,r) = \sum_{n,m =1}^{\infty} { (m+n)^{-p}
 m^{-q}  n^{-r}}$. By decomposing the summands
for fixed $n$ into partial sums with $m$-dependent
 singularities and
 $n$-dependent coefficients, one derives the following
identities
\eqn\AA{\eqalign{
I(1,2,1) &=   I(1,0,3) + I(0,2,2) - I(0,1,3) \cr
I(2,1,1) &= - I(2,0,2) - I(1,0,3) + I(0,1,3) \cr
I(3,1,0) &= - I(3,0,1) - I(2,0,2) - I(1,0,3) + I(0,1,3)  \cr
I(1,3,0) &= - I(1,0,3) + I(0,3,1) - I(0,2,2) + I(0,1,3). \cr}}
 (From $I(2,2,0)$ one obtains the same result as from $I(1,3,0)$.)
{}From these relations, one finds
$I(1,2,1)$, $I(2,1,1)+ I(2,2,0)$
and $2I(3,1,0) + I(2,2,0)$,
which agree with \A.

To obtain also $I(2,1,1)$, A. Martin gave us the following derivation
\eqn\AAA{\eqalign{ &I(2,1,1) =
\sum_{n,m =1}^{\infty} {1\over { (m+n)^2 mn}}=
\sum_{n,m =1}^{\infty} \int_0^1 {dx \over x}
\int_0^x {dy\over y}{y^m\over m} {y^n\over n} \cr
& = \int_0^1 {dx \over x} \int_0^x {dy\over y}[-\ln (1-y)]^2
= \int_0^1 {dy\over y} \ln^2 (1-y) \int_y^1 {dx\over x} = - \int_0^1
{dx \over x } \ln^2(1-x) \ln x. \cr}
}
Next we replace $\ln (1-x) $ by $ {{(1-x)^\epsilon -1} \over \epsilon}$,
but for $\ln x$  we use $ {{ x^{2\epsilon} - x^\epsilon} \over \epsilon}$
to cancel the ${1\over x}$ singularity.
One then
obtains  a series of beta functions $B(x,y) =\int_0^1 dt\
t^{x-1}
 (1-t)^{y-1}
= {\Gamma(x) \Gamma(y) \over {\Gamma(x+y)}}$, $\Gamma$ being the
gamma function:
\eqn\B{I(2,1,1) =-{1\over {\epsilon^3}} \biggl [ B(2\epsilon, 1+2\epsilon)
- B(\epsilon,1+2\epsilon) -2B(2\epsilon,1+\epsilon) +2B(\epsilon,1+\epsilon)
+{1\over {2\epsilon}} - {1\over \epsilon} \biggr ]. }
Next we use $\Gamma(2\epsilon)= {\Gamma(1+2\epsilon) \over {2 \epsilon}}$
and obtain
\eqn\BB{I(2,1,1)
= -{1\over {2\epsilon^4}}\biggl [ \biggl (
{\Gamma^2(1+2\epsilon) \over {\Gamma(1+4 \epsilon)}} - 1 \biggr) +4
\biggl (
{\Gamma^2(1+\epsilon) \over {\Gamma(1+2 \epsilon)}}-
{\Gamma(1+\epsilon) \Gamma(1+2\epsilon) \over {\Gamma(1+3 \epsilon)}}
\biggr ) \biggr]. }
Writing $\Gamma (1+ z)$ as $\exp ( z \psi(1) +{1\over 2} z^2 \psi'(1) +
\dots)$ where $\psi(z) = {\Gamma'(z) \over {\Gamma(z)}}$
is the psi (digamma) function, given also by $\psi'(z) =
\sum_{n =0}^{\infty}  { (n + z)^{-2}}$, we see
that  all $\psi(1)$
and all even derivatives of $\psi(z)$ at $z=1$ cancel.
The $2n+1$ derivatives of $\psi(z)$ at $z=1$
are proportional to $\zeta(2n)$,
which are known:
$\psi'(1) = \sum_{n=1}^{\infty} n^{-2} = {\pi^2 \over 6}$,
$\psi'''(1) = 6\sum_{n=1}^{\infty} n^{-4} = {\pi^4 \over {15}}$.
Substitution of these expansions into $I(2,1,1)$ then reveals
that all $\epsilon$ poles cancel, and the finite result is
$I(2,1,1)= {\pi^4 \over {180}}$.
\smallskip

We thank A. Schellekens for discussions, and expecially
A. Martin for a derivation of $I(2,1,1)$.

\appendix{B}
\indent
In this  appendix we comment on some problems
that are one encountered in the
operator approach to anomalies. We start from $N$ operators
 $\hat A_1,\dots, \hat A_N$
depending on  canonical variables
 $ \hat a = {( \hat q + i \hat p) \over {\sqrt 2}} $ and
$ \hat a^{\dagger} = {( \hat q - i \hat p) \over {\sqrt 2}} $,
satisfying $[ \hat a, \hat a^\dagger] = 1$. [If the Hamiltonian
is $H= {1\over {2m}} P^2 +{1\over 2} m \omega^2 Q^2$, then
$q= (m \omega)^{1\over 2} Q$ and $p= (m \omega)^{-{1\over2}} P$,
so that $dPdQ=dpdq$ and $H=( \hat a^\dagger \hat a +{1\over 2})
\omega$].
Using the so-called  holomorphic representation \ref\fad{L. Faddeev,
in \lq\lq Methods in Field Theory\rq\rq,
eds. R. Balian and J. Zinn-Justin (North Holland,
 Amsterdam, 1976)\semi
C. Itzykson and J.B. Zuber, \lq\lq Quantum Field Theory\rq\rq
(Mc Graw-Hill, New York, 1980).
}, \ref\agl{L. Alvarez-Gaum\'e, in  \lq\lq
Supersymmetry\rq\rq,
Nato-ISA series vol.N125, eds. K. Dietz, R. Flume, G.C. von Gehlen
and V. Rittenberg (Plenum Press, New York, 1984).},
the trace of the product of the $\hat A_i$ can be written as a  multiple
integral (where $dw d \bar w = idpdq $, etc.)
\eqn\eBu{ {\rm Tr} ( \hat A_1,\dots, \hat A_N) =
\int  {dw d \bar w \over { 2 \pi i}}
\prod_{j=1}^{N-1}
{ d \bar v_j dv_j \over { 2 \pi i}}
\exp \biggl ( - w \bar w -\sum_{i=1}^{N-1}
\bar v_i v_i \biggr )
A_1(\bar w,v_1) \dots A_N (\bar v_{N-1},w).}
The kernels $A(\bar z, z)$ are equal to  $(\exp \bar z z )
K(\bar z, z)  $,
where $ K(\bar  z,z)= \sum_{m,n \geq 0} \bar z^m K_{mn} z^n $ if
$ \hat A = \sum_{m,n \geq 0} \hat a^{\dagger m} K_{mn}  \hat a^n$.
Note that this last expression has to be normal ordered with
respect to
$\hat a^\dagger$ and $\hat a$.

Consider now $\exp (-\beta \hat H )= \bigl ( \exp (- \epsilon \hat H)
\bigr)^N $,
where $N \epsilon = \beta$.
Let $\exp (- \epsilon \hat H (\hat a^\dagger, \hat a ) )= $
$ : \exp (-\epsilon
\hat h (\hat a^\dagger, \hat a ) ): $
where $\hat h$ will in general depend on
$\epsilon$, then the trace can be written as
\eqn\eBd{\eqalign{ {\rm Tr} e^{-\beta \hat H} &=
\int  {dw d \bar w \over { 2 \pi i}}
\prod_{j=1}^{N-1}
{ d \bar v_j dv_j \over { 2 \pi i}}
\exp ( - L ) \cr
 - L &= -\bar w ( w - v_1) - \bar v_1 ( v_1 - v_2) + \dots - \bar v_{N-1}
(v_{N-1} - w)\cr
& \ \ \  - \epsilon [ h(\bar w, v_1) + h(\bar v_1, v_2) + \dots
 h(\bar v_{N-1}, w) ]. \cr}}
By identifying $ w=v_N $
and $\bar w=\bar v_0$, this result can be written in a suggestive form
as a path integral with  periodic boundary conditions (PBC)
\eqn\eBt{ {\rm Tr}\ {\rm e}^{-\beta \hat H} =
\int_{PBC} \prod_{\tau} { dv(\tau) d \bar v(\tau) \over { 2 \pi i}}\
\exp \biggl (\int_0^\beta
 d\tau \bigl ( \bar v {d v\over {d\tau}} - h(\bar v,v)
\bigr ) \biggr), }
but the exact meaning (exact even for finite $N$) is given by \eBd.

Given the operator $\hat H$ (the regulators in our case), it is very difficult
to find $h(\bar z,z)$ in closed form. One has to  expand $\exp (
 - \epsilon \hat H )$,
commute all $\hat a^\dagger$ to the left and $\hat a$ to the right,
replace  $\hat a^\dagger$ by $z$ and $\hat a$ by $z$, and finally
re-exponentiate. For the harmonic oscillator this has been done but not,
to our knowledge, for non-linear sigma models.
For the harmonic oscillator with $\hat H = (\hat a^\dagger \hat a +{1\over 2})
\omega$ one obtains
\ref\louise{W.H. Louisell,
\lq\lq Radiation and Noise in Quantum Electronics\rq\rq
(Mc Graw-Hill, New
York, 1964)\semi
J. Katriel, J. Phys. A16 (1983) 4171.}
\eqn\eBtb{\exp \bigl (-\epsilon h(\bar z,z)\bigr )
= \exp \biggl [ -{1\over 2} \epsilon \omega - \exp ( 1 - {\rm e}^{-\epsilon
\omega}) \bar z z \biggr ]. }

To obtain the configuration space equivalent of \eBd, we must integrate out
the $N$ momentum variables $p,p_1,\dots,p_{N-1}$. In  the
continuum limit \eBt\ we find from the term $\int_0^\beta
d \tau \bar v { d v \over
{d\tau}}$, upon re-discretization,
a sum $\sum_{j=0}^{n-1} {1\over 2} ( q_j -ip_j ) ( \dot q_j +i \dot p_j)
\epsilon$.
Owing  to PBC, this can be rewritten as a sum of $ {1\over 4} {d \over
{d \tau}} ( q_j^2 + p_j^2 )\epsilon - ip_j \dot q_j$,
and if $h(p,q)$ is quadratic in $p$ [so $h(p,q) =
{1\over 2} g^{ij} (q) p_i p_j +V(q)$]
one obtains the Euclidean action
$L = \int_0^\beta d\tau ({1\over 2} g_{ij} \dot q^i \dot q^j + V) $
together with the measure $\sqrt g$ discussed in sect. 1.

However, if one starts from the exact result in \eBd, the computation becomes
considerably
more complicated. Consider first the harmonic oscillator. One obtains
then in the exponent an expression of the form $ - {1\over 2}
\sum ( q_j -i  p_j ) M_{jk} (q_k + i p_k)  \epsilon - {1\over 2}
N \epsilon \omega$, where
the non-symmetric matrix $M$ has entries
$+1$
along the diagonal, and $ - {\rm e}^{- \epsilon \omega}$
just above the diagonal and in the lower-left position.
The integration  over the $N$ variables $p_j$ yields
\eqn\eBq{ (2\pi)^{N\over 2} ({\rm det} M_S)^{-{1\over 2}}
\int_{PBC} \prod_{j=0}^{N-1} {d q_j\over {2 \pi}}  \exp
\biggl ( -{1\over 2} q ( M_S - M_A M_S^{-1}
M_A ) q \biggr ) }
and further integration over the $q$ variables yields the expected result
 ${\rm det} M = 1 - \exp ( - \epsilon \omega N)$.
Together  with the factor $( \exp -{1\over 2} \epsilon \omega)^N$,
one finds
$  ( 2 \sinh  {\beta \omega \over 2}  )^{-1}$,
which is indeed  the partition function for the harmonic oscillator.
To see ${\rm det} M$  coming out, note that
\eqn\eBc{ {\rm det} M_S \ {\rm det} ( M_S - M_A M_S^{-1}
M_A ) = {\rm det} M_S (M_S- M_A ) M_S^{-1} (M_S +M_A),}
where $M_S = {1\over 2} ( M + M^T)$ and
$M_A = {1\over 2} ( M - M^T)$.
For a system
more general  than the harmonic oscillator, one would have to
compute ${\rm det } M_S$ in order to obtain the exact corresponding
configuration space path integral, and this is  in general difficult.
(Even for the harmonic oscillator, we have not found ${\rm det} M_S$.
In \schul\  the determinant is given of a symmetric matrix, which is equal to
our  $M_S$ except that the lower left entry and upper right entry are zero.)

For the
chiral anomaly, one must evaluate $ {\rm Tr}  \exp (i \theta \hat F)
\exp (-\beta \hat H)$, where $\hat F= \hat c^\dagger \hat c $ is the
fermion number operator.
The integral kernel for $ \hat A_1 = \exp (i \theta \hat F)$ can be found in
closed form, because
the  variables $c$ and $\bar c$ are anticommuting:
$ A_1(\bar c,c) = \exp ( {\rm e}^{i\theta} \bar c c)$.
The path integral corresponding to \eBt\ is obtained by replacing the first
factor $\exp (- \epsilon \hat H)$ by  $ \exp (i\theta \hat F)$,
and one finds
\eqn\eBs{\eqalign{ {\rm Tr}\ {\rm e}^{i\theta \hat F}
{\rm e}^{-\beta \hat H} &=
\int  {d\xi d \bar \xi \over { 2 \pi i}}
\prod_{j=1}^{N-1}
{ d \bar \xi_j d\xi_j \over { 2 \pi i}}
\exp ( - L ) \cr
 - L &=  \bar \xi_{N-1} (\xi - \xi_{N-1})
 + \dots + \bar \xi_1 ( \xi_2 - \xi_1)
 + \bar
\xi {\rm e}^{i \theta} (\xi_1 + {\rm e}^{-i \theta} \xi ) \cr
& \ \ \ - \epsilon [ h(\bar \xi_1, \xi_2)
+ \dots
 h(\bar \xi_{N-1}, \xi) ]. \cr}}
We have taken care to define \eBu\ such that it also holds for anticommuting
variables, as in \eBs.
By identifying $\xi_N = \xi$ and $ \xi_0 = - {\rm e}^{-i\theta} \xi$
we obtain  a path integral as in \eBt, but now with antiperiodic boundary
conditions ABC for $\theta = 0$, and PBC for $\theta = \pi$.
Hence, for the trace anomaly we would need fermions with ABC, see sect. 1.

Finally, we consider  the ghosts $\hat c^{i \dagger}$ and $\hat c_i$
for internal symmetry, discussed in sect. 1. We have found the projection
operator $P_1$ onto the one-particle states $ \hat c^{i\dagger} |0\! >$
in closed form. It  is given by
\eqn\eBl{ P_1 = : \exp( \hat c^{i\dagger} \hat c_i ) : - 1.}
Written out, $ P_1 = \hat c^{i\dagger} \hat c_i
+ {1\over 2} \hat c^{i\dagger}  \hat c^{j\dagger} \hat c_j \hat c_i + \dots$,
and it is clear that $P_1 |0\! > = 0$ and $P_1 \hat c^{i\dagger} |0\! > =
\hat c^{i\dagger} |0\! >$.
One can check that $P_1$ annihilates the states with  $2, \dots, M$ particles
(where $j=1, \dots,M$).
It follows that the analysis is very similar to that of the chiral anomaly:
one must only replace the integral kernel
$\exp ( {\rm e}^{i \theta} \bar c c )$ by $\exp (\bar c c) - 1$.
However, we have not been able to interpret this result in terms of suitable
boundary conditions.

Our conclusion is that the operator approach, which starts from the
quantum Hamiltonian,
is much more complicated than the Feynman approach, which produces
the Hamiltonian
at the end. For this reason we have followed in  the main text
the Feynman approach (but we have not made the error of omitting
quantum fluctuations from the path integral
\ref\cheng{K.S. Cheng, J. Math. Phys. 13 (1972) 1723.}).
The approach of the authors in \ag\
 was based on an operator approach, but they made various  approximations:
they assumed that  $h=H$, and used  \eBt\ instead of \eBd,
to obtain the Euclidean action.
For the chiral anomalies, their approximate path integral yields the
correct result because  chiral anomalies depend on very few details of
the path integral (they are topological).
For the trace anomaly these approximations would be incorrect, as we
discussed in the introduction.

\listrefs
\end

\appendix{C}

For completeness, we review here the proof of the transformation
properties of path ordered exponentials used above. Given a path
$q^\m(\tau)$ joining $q^\m(0)=y^\m$ to $q^\m(\beta)= x^\m$
in a time lapse $\beta$,
we construct the group element $G(x,y) \equiv
G(\beta) $ associated with such a
path by solving the differential equation
\eqn\diff{ \eqalign{
&{d \over{ d \tau}} G(\tau) = - \dot q^\mu A_\mu  G(\tau) \cr
&G(0)=1\cr}}
We first turn such a differential equation
into an integral equation
\eqn\integral{ G(\beta) = 1 - \int_0^\beta d\tau \  \dot q^\m A_\m G(\tau)}
which is solved by the path ordered exponential
\eqn\po{ G(\beta) = T \e^{- \int_0^\beta d\tau \  \dot q^\m A_\m}}
To check its transformation properties under the gauge transformation
\gt,
we look at the corresponding equation with $A_\m'$
\eqn\dif{
{d \over{ d \tau}} G'(\tau) = - \dot q^\mu A_\mu'  G'(\tau)=
- \dot q^\mu U A_\mu U^{-1} G'(\tau) - U \dot U^{-1} G'(\tau)}
which can be recasted in a form identical to equation \diff
\eqn\difff{
{d \over{ d \tau}} (U^{-1}G') = - \dot q^\mu A_\mu  U^{-1}G' }
Because this is the same equation as for $G$, we can identify
the two solutions once the same boundary conditions are imposed
since the solution is then unique. Thus
\eqn\result{ G' (\beta) = U(\beta) G(\beta) U^{-1}(0)}
This is the gauge transformation of the path ordered exponential
that we have used above and which we rewrite below in  more detailed
notation
\eqn\popo{T \e^{- \int_0^\beta d\tau \  \dot q^\m A_\m'}
= U\bigl (q(\beta)\bigr)
\bigl( T
\e^{- \int_0^\beta d\tau \  \dot q^\m A_\m}
\bigr) U^{-1}\bigl ( q(0) \bigr) }
A similar proof goes through if a scalar potential $V$
transforming in the adjoint representation is included in
the exponent.

--------------------------------

 However, note that the
 gauge fixing term in \agfgf\ naively destroys  the Weyl invariance.
In fact, even by extending the Weyl transformations in a natural way
 on the $\Pi,B,C$ fields
\eqn\2{\eqalign{ &\Pi \to \Pi' = \Omega^{-1} \Pi \cr
& B \to  B' = \Omega^{-1} B \cr
& C \to C' = C \cr}}
the action is not invariant and there is no way of fixing the above
transformation rules to achieve invariance.
Under an infinitesimal Weyl transformation, when $\Omega = e^\s =
1+ \s+\dots$, the action changes by
\eqn\change{\eqalign{\delta S_1^{gf} &= \int d^4x {\sqrt g}\
{\rm Tr} \biggl \lbrack \Pi (\Delta \Gamma^\m) A_\m +
B (\Delta \Gamma^\m) (\pa_\m C + [ A_\m , C ] )\biggr \rbrack \cr
&=
\delta_{^{BRST}} \int d^4x {\sqrt g}\  {\rm Tr}  \biggl \lbrack B
(\Delta \Gamma^\m)
A_\m \biggr \rbrack\cr}}
with $ (\Delta \Gamma^\m) \equiv  g^{\m\n} \pa_\n \s$.
In
 the last line  we have made use
of the anticommuting BRST operator defined by
\eqn\BRST{ \eqalign{&\delta_{^{BRST}} A_\m = \pa_\m C +[A_\m , C] \cr
&\delta_{^{BRST}} C = -{1\over 2} \{ C,C \} \cr
&\delta_{^{BRST}} B = \Pi \cr
&\delta_{^{BRST}} \Pi = 0 \cr}}
which shows that  the Weyl variation of the action is BRST exact.
Because of this, the breaking of the Weyl symmetry has no
effect on BRST invariant states, which are then Weyl covariant
as well.

----------------

 (Since time-ordering is crucial for the calculations we give a
derivation of this result. One may check, for example, that
$\int_{-1}^{0}dt \int_{-1}^{t}dt'(\dot q^{\m}A_{\m})(t)
\ (\dot q^{\m}A_{\m})(t')$ varies under $\d A_{\m}=\pa_{\m}\phi$ into
$\phi(0) \int_{-1}^{0} \dot q^{\m}A_{\m}dt - \int_{-1}^{0} \dot q^{\m}A_{\m}
dt \phi(-1)$ which is the desired result. More generally, since the
time-ordered path-integral as a product of Riemann factors
$[1+\epsilon \dot q^{\m}A_{\m}]$ satisfies the same boundary conditions
and the same differential equation, ${d \over dt}U=\dot q^{\m}A_{\m}U$, as
the expression with $\int_{-1}^{0}dt\int_{-1}^{t}dt_1  \cdots $, gauge
covariance of the latter expression is manifest.)

---------------

($ T \exp (\int_{t_i}^{t_f} \dot q^\m A_\m') =
U\bigl ( q(t_f) \bigr )
T \exp (\int_{t_i}^{t_f} \dot q^\m A_\m)
U^{-1}\bigl (q(t_i)\bigr )$).

 and we can start computing the various
diagrams. It is appropriate to recall  that the measure
\pim\ normalizes
the leading Gaussian integral to  $A =  { (2 \pi \b)^{-{d\over 2}}} $.
Because we now no longer
integrate over $z_i^\m$,
this implies that to  compute  the trace
\ty\ for $d=2$ we need

\ref\louise{W.H. Louisell,
\lq\lq Radiation and Noise in Quantum Electronics\rq\rq, Mc Graw-Hill, New
York, 1964.},

We should remember ref. \ref\fII{K. Fujikawa, S. Ojima and S. Yajima,
Phys. Rev. D34 (1986) 3223}

We have used quantum mechanics to compute trace anomalies. Of
interest is the derivation itself. (The trace anomalies
in $d=2$ and $d=4$ are, of  course, well-known, and  our results agree
 with the literature \duff.
We have not obtained a closed formula for the trace anomalies
in any $d$, similar to the result of ref. \ag\  for
the chiral anomaly based on the \lq\lq Dirac genus\rq\rq,
but a starting point might be our formula \kak).
The approach based on quantum mechanics is simpler than
other methods and it is an excellent tool for computing anomalies.
However, for the trace anomalies, certain fundamental and interesting
aspects of  path integrals which could be neglected in the case of
chiral anomalies, had to be taken into account. As already observed in ref.
\new, this makes the computation of the trace anomalies
more complicated than that of chiral anomalies.